\documentclass[10pt,technote]{IEEEtran}

\usepackage{amsmath,epsfig,amssymb,verbatim}
\usepackage{cite}
\usepackage{graphicx}
\usepackage{bm,bbm}
\usepackage{xcolor}
\usepackage{subfigure}
\usepackage{algorithm}
\usepackage{algpseudocode}
\usepackage{stackengine}
\usepackage{cleveref}
\usepackage{multicol}
\usepackage{multirow}

\newtheorem{Lemma}{Lemma}

\begin{document}
\title{Low Complexity Artificial Noise Aided Beam Focusing Design in Near-Field Terahertz Communications}

\author{Zhifeng Tang, \IEEEmembership{Member, IEEE,} Nan Yang, \IEEEmembership{Senior Member, IEEE,} Xiangyun Zhou, \IEEEmembership{Fellow, IEEE,} Salman Durrani, \IEEEmembership{Senior Member, IEEE,} Markku Juntti, \IEEEmembership{Fellow, IEEE,} and Josep Miquel Jornet, \IEEEmembership{Fellow, IEEE}


\thanks{This work was supported by the Australian Research Council Discovery Project (DP230100878). This work has been submitted to the IEEE Transactions on Vehicular Technology for possible publication. Copyright may be transferred without notice, after which this version may no longer be accessible.}
\thanks{Z. Tang, N. Yang, X. Zhou, and S. Durrani are with the School of Engineering, Australian National University, Canberra, ACT 2601, Australia (Emails: \{zhifeng.tang, nan.yang, xiangyun.zhou, salman.durrani\}@anu.edu.au).}
\thanks{M. Juntti is with the Centre for Wireless Communications, University of Oulu, Oulu 90014, Finland (Email: markku.juntti@oulu.fi).}
\thanks{J. M. Jornet is with the Department of Electrical and Computer Engineering, Northeastern University, Boston, MA 02120, USA (Email: j.jornet@northeastern.edu).}}

\maketitle

\begin{abstract}
In this correspondence, we develop a novel low-complexity artificial noise (AN) aided beam focusing scheme in a near-field terahertz wiretap communication system. In this system, the base station (BS) equipped with a large-scale array transmits signals to a legitimate user, while mitigating information leakage to an eavesdropper. We formulate an optimization problem to maximize the secrecy rate achieved at the legitimate user and solve it by designing the optimal beam focusing and power allocation. 
Numerical results demonstrate the significant performance improvement achieved by the proposed AN aided beam focusing scheme, especially when the eavesdropper is located closer to the BS than the legitimate user.
\end{abstract}


\vspace{-3mm}

\section{Introduction}

With the growing demand for ultra-fast and massive data transmission, terahertz (THz) communications has emerged as a game-changing technology for the sixth-generation (6G) wireless networks \cite{6GNet2020}. In particular, THz communications operates in the THz frequency range, i.e., 0.1--10 THz, offering extensive spectrum and enabling picosecond-level symbol duration \cite{THzComMCS2024}. With ultrabroad bandwidth and ultra-high data rate, THz communications unlocks rich opportunities for the Internet of Things (IoT) applications, such as ultra-fast data transfer in wireless data centers \cite{Tang2025TCom}. 


It is noted that the Rayleigh distance in THz communication systems can be significantly larger than that at lower frequencies, due to the short wavelength of THz propagation and the use of large-scale antenna arrays in such systems, which results in a large near-field region \cite{Bodet2024}. This large region offers enhanced physical layer security by confining the signal to a smaller spatial area, thus limiting the potential for interception by unauthorized users. However, a high risk of information leakage to nearby eavesdroppers still remains \cite{Ferreira2024access}.
In light of this, \cite{Zhang2024tvt} proposed a near-field point-to-point transmission framework and a beamformer to optimize the secrecy rate. Also, \cite{Zhang2024tvt} revealed that the near-field secrecy performance primarily relies on the distance disparity between the legitimate user and the eavesdropper. In \cite{Ali2024coml}, an extended multi-user near field communication system was introduced, where hybrid beamforming was developed based on successive convex approximation and optimization to maximize the minimum secrecy rate among all legitimate users. A wavefront hopping scheme was proposed in \cite{Petrov2024tcom} to decrease the probability of message eavesdropping. {In addition, the null-space artificial noise (AN) of the legitimate channel was proposed in \cite{Chen2024tvt} to improve the secrecy rate. However, this approach becomes less effective when the channel steering vectors of the legitimate user and the eavesdropper are highly correlated. In this scenario, the null spaces of their channel steering vectors significantly overlap, limiting the ability to direct interference toward the eavesdropper and thereby degrading secrecy performance.} Although the aforementioned studies have presented effective methods to improve the secrecy performance in near-field communication systems, these methods introduced high computational complexity, especially when THz communication systems deploy extremely large-scale arrays, which brings huge challenges for practical implementation. Very recently, an AN aided beam focusing scheme was proposed in \cite{zhang2024performanceanalysislowcomplexitybeamforming}, {where the authors designed signal beam focusing toward the legitimate users and AN beam focusing toward the eavesdropper. This approach effectively reduces computational complexity while providing moderate improvements in secrecy performance.}
To further improve the secrecy performance with low computational complexity, we will devise a novel AN aided beam focusing scheme, motivating this work.

\begin{figure}[t]
\centering
\includegraphics[width=0.75\columnwidth]{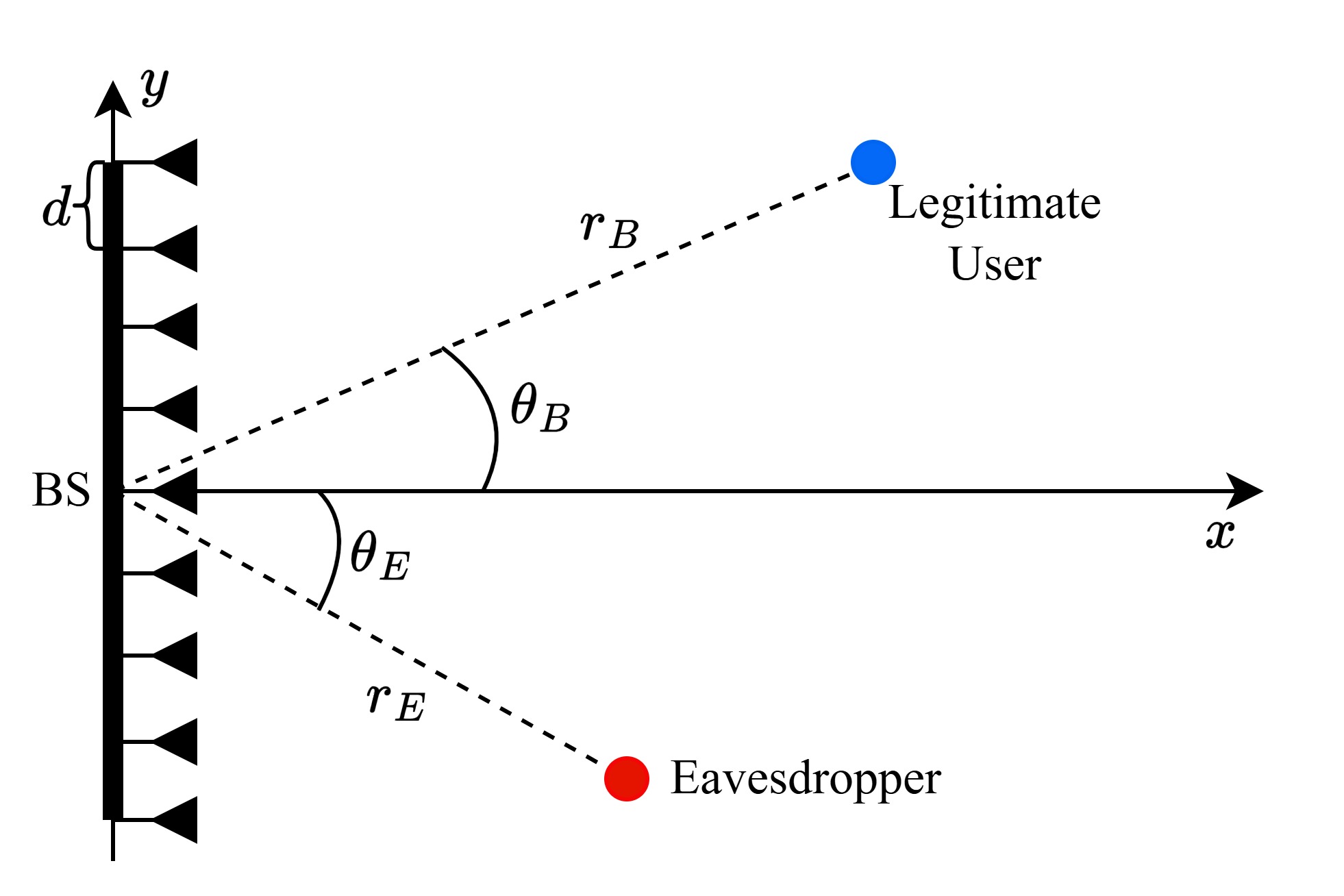}
\vspace{-1em}
\caption{Illustration of our considered near-field communication system.}\label{fig:system_model}
\vspace{-2em}
\end{figure}

In this letter, we consider the downlink in a near-field THz wiretap communication system, where a base station (BS) transmits signals to a legitimate user while an eavesdropper exists in the system to wiretap the transmission. 
Inspired by AN aided beam focusing design \cite{zhang2024performanceanalysislowcomplexitybeamforming}, we develop a novel low-complexity AN aided beam focusing scheme to improve the secrecy performance. {In particular, aided by approximation techniques, we first decouple the secrecy rate optimization problem into separate signal and AN beam focusing subproblems and solve each subproblem by a low-complexity one-dimensional search method to find the optimal signal and AN beam focusing points, while considering their combined impact on both the legitimate user and the eavesdropper. }
We then derive the closed-form expression for the optimal power allocation ratio on AN transmission to maximize the secrecy rate. Through numerical results, we demonstrate that our proposed scheme achieves profound secrecy rate improvement compared to the existing ones.

\section{System Model}\label{Sec:System}

We consider the downlink of a near-field THz wiretap communication system, as depicted in Fig. \ref{fig:system_model}. In this system, a BS transmits signals to a legitimate user, while one eavesdropper exists to wiretap this signal\footnote{{While we consider a single-user and single-eavesdropper configuration in this work to highlight the fundamental design principles, the proposed scheme can be extended to multi-user and multiple-eavesdropper scenarios. In such scenarios, user–eavesdropper pairs are formed based on the correlations between their channel steering vectors, where each legitimate user is paired with the eavesdropper having the highest correlation. The secrecy design problem can then be decoupled into parallel sub-problems for each pair, which can be solved independently. Here, we note that different legitimate users may be paired with the same eavesdropper. In this case, this eavesdropper experiences a higher AN than that considered in each individual sub-problem, where each sub-problem characterizes the worst-case scenario for the corresponding legitimate user. Subsequently, joint power allocation across all user-eavesdropper pairs is optimized to meet individual secrecy requirements and system power constraints.}}. Specifically, the BS employs a uniform linear large-scale array with $N$ antenna elements, while both the legitimate user and the eavesdropper are standard low-cost receivers (e.g., sensors in THz IoT applications) equipped with a single antenna \cite{Chen2024tvt,zhang2024performanceanalysislowcomplexitybeamforming,Zhang2024tsp}. We assume that the legitimate user and the eavesdropper are located in the near-field region, i.e., at distances smaller than the Rayleigh distance, given by $d_R = {2D^2}/{\lambda}$, but greater than the Fresnel distance, given by $d_F = 0.62\sqrt{{D^3}/{\lambda}}$, where $D$ is the antenna aperture size of the BS and $\lambda$ is the wavelength. {Furthermore, we assume that both the legitimate user and the eavesdropper are active users in this system, while the eavesdropper is unauthorized to access the signal transmitted to the legitimate user in the considered transmission. Therefore, the perfect channel state information of the legitimate user and the eavesdropper is available at the BS via effective near-field channel estimation and beam training methods, such as those in \cite{liu2024sensingenhancedchannelestimationnearfield,Zhang2022wcl}.}

\subsection{Channel Model}

Without loss of generality, we assume that the large-scale array is placed along the $y$-axis, while the center of the large-scale array is located at the origin, i.e., $(0,0)$. Accordingly, the $n$th antenna element, $n\in\{1,\cdots,N\}$, has the coordinate $\left(0,\delta_n d\right)$, where $\delta_n=n-\frac{N-1}{2}$ and $d$ represents the spacing between adjacent antenna elements. Moreover, the coordinates of the legitimate user and the eavesdropper are given by $L_B=\left(r_B\cos\theta_B,r_B\sin\theta_B\right)$ and $L_E = \left(r_E\cos\theta_E,r_E\sin\theta_E\right)$, respectively, where $r_{B}$ and $r_E$ denote the radial distance of the legitimate user and the eavesdropper from the origin, respectively, and $\theta_{B}$ and $\theta_{E}$ denote the spatial angle from the $x$-axis to the legitimate user and the eavesdropper, respectively. As such, the distances from the $n$th antenna element of the large-scale array to the legitimate user and the eavesdropper are expressed as
\begin{align}
r_{n,B} = \sqrt{r_{B}^2+(\delta_n d)^2-2r_B\delta_n d\sin\theta_B}
\end{align}
and
\begin{align}
r_{n,E} = \sqrt{r_{E}^2+(\delta_n d)^2-2r_E\delta_n d\sin\theta_E},
\end{align}
respectively.

{In near-field communications, the multipath channel model is usually adopted, which consists of a line-of-sight (LoS) path and several non-LoS (NLoS) paths \cite{Bodet2024}. Moreover, due to the identical channel gains but non-uniform phases from different antennas, the near-field channel from the BS to the legitimate user and the eavesdropper are expressed as}
\begin{align}
\mathbf{h}_{B}^{H} = h_{B,\mathrm{LoS}}\mathbf{h}^{H}(r_B,\theta_B) + \sum\limits_{m=1}^{M_B}\mathbf{h}_{B,\mathrm{NLoS},m}^{H}
\end{align}
{and}
\begin{align}
\mathbf{h}_{E}^{H} = h_{E,\mathrm{LoS}}\mathbf{h}^{H}(r_E,\theta_E) + \sum\limits_{m=1}^{M_E}\mathbf{h}_{E,\mathrm{NLoS},m}^{H},
\end{align}
{respectively, where $h_{B,\mathrm{LoS}} = \frac{\lambda}{4\pi  r_{B}}\exp\left(-\frac{1}{2}K(f)r_B\right)$ and $h_{E,\mathrm{LoS}} = \frac{\lambda}{4\pi  r_{E}}\exp\left(-\frac{1}{2}K(f)r_E\right)$ denote the large-scale path loss coefficient from the origin to the legitimate user and the eavesdropper, respectively, $K(f)$ is the molecular absorption coefficient at frequency $f$, and $\mathbf{h}(r_B,\theta_B)$ and $\mathbf{h}(r_E,\theta_E)$ are the channel steering vectors, given by}
\begin{align}
\mathbf{h}^H(r_B,\theta_B) = \left[e^{-j2\pi r_{1,B}},\cdots,e^{-j2\pi r_{N,B}}\right]
\end{align}
{and}
\begin{align}
\mathbf{h}^H(r_E,\theta_E) = \left[e^{-j2\pi r_{1,E}},\cdots,e^{-j2\pi r_{N,E}}\right],
\end{align}
{respectively. Furthermore, $M_B$ and $M_E$ denote the very limited number of scatterers associated with the legitimate user and eavesdropper, respectively. The vectors $\mathbf{h}_{B,\mathrm{NLoS},m}^{H}$ and $\mathbf{h}_{E,\mathrm{NLoS},m}^{H}$ represent the NLoS channel from the BS to the legitimate user and the eavesdropper, respectively, arising from reflections at the $m$th scatter, as defined in \cite[Eq. (132)]{Liu2023ojcom}. Due to the severe path loss induced negligible power of the NLoS channel in THz communications \cite{Shafie2021JSAC}, the channels from the BS to the legitimate user and the eavesdropper can be approximated as $\mathbf{h}_{B}^{H} \approx h_{B,\mathrm{LoS}}\mathbf{h}^{H}(r_B,\theta_B)$ and $\mathbf{h}_{E}^{H} \approx h_{E,\mathrm{LoS}}\mathbf{h}^{H}(r_E,\theta_E)$, respectively.}

\subsection{Signal Model}

We assume that the BS transmits the information-carrying signal along with the AN to serve the legitimate user while disrupting the eavesdropper. We denote $\mathbf{x}\in \mathbb{C}^N$ as the signal vector transmitted from the BS to the legitimate user, given by $\mathbf{x} = \mathbf{w_s}s + \mathbf{w_z}z$, where $\mathbf{w_s}\in \mathbb{C}^N$ denotes the beam focusing vector of the signal, $s$ denotes the information-carrying signal for the legitimate user, $\mathbf{w_z}\in \mathbb{C}^N$ denotes AN vector, and $z$ denotes the AN. Hence, the received signals at the legitimate user and the eavesdropper are given by
\begin{align}\label{eq:signalB}
y_B =  \mathbf{h}_{B}^{H} \left(\mathbf{w_s}s + \mathbf{w_z}z\right)+n_{B}
\end{align}
and
\begin{align}\label{eq:signalE}
y_E =  \mathbf{h}_{E}^{H} \left(\mathbf{w_s}s + \mathbf{w_z}z\right)+n_{E},
\end{align}
respectively, where $n_B\sim \mathcal{CN}(0,\sigma_B^2)$ and $n_E\sim \mathcal{CN}(0,\sigma_E^2)$ denote the additive white Gaussian noise (AWGN) at the legitimate user and the eavesdropper, respectively. Here, we assume that the legitimate user and the eavesdropper have the same noise power, i.e., $\sigma_B^2=\sigma_E^2=\sigma^2$. {According to \eqref{eq:signalB} and \eqref{eq:signalE}, the achievable rates of the legitimate user and the eavesdropper are given by}
\begin{align}
R_B = \log_2\left(1+\frac{|\mathbf{h}_{B}^{H} \mathbf{w_s}|^2}{|\mathbf{h}_{B}^{H}\mathbf{w_z}|^2+\sigma^2}\right)
\end{align}
{and }
\begin{align}
R_E = \log_2\left(1+\frac{|\mathbf{h}_{E}^{H} \mathbf{w_s}|^2}{|\mathbf{h}_{E}^{H}\mathbf{w_z}|^2+\sigma^2}\right),
\end{align}
{respectively. Therefore, the achievable secrecy rate of the legitimate user is given by $R_S = \left(R_B-R_E\right)^+$, where $(x)^+=\max(x,0)$. }

\section{Problem Formulation and Proposed Solution}\label{Sec:Coverage}

In our considered system, our design aims to maximize the secrecy rate by optimizing the beam focusing vector and AN, i.e., $ \mathbf{w_s}$ and $\mathbf{w_z}$. To achieve this aim, we formulate the optimization problem as
\begin{subequations}
\begin{align}
\mathbf{P1}:\quad &\max\limits_{\mathbf{w_s},\mathbf{w_z}}\  R_S\\
&\mathrm{s.t.}\  ||\mathbf{w_s}||^2+||\mathbf{w_z}||^2\leq P,\label{eq:probconst}
\end{align}
\end{subequations}
where \eqref{eq:probconst} denotes the transmit power constraint of the BS with the maximum transmit power $P$.

We next develop a low-complexity beam focusing scheme to transmit the information-carrying signal and AN for secrecy range maximization. We assume that the signal and AN beam focusing points, denoted by $Q_S$ and $Q_A$, have the coordinates $L_S=(r_S\cos\theta_S,r_S\sin\theta_S)$ and $L_A=(r_A\cos\theta_A,r_A\sin\theta_A)$, respectively. Thus, the near-field channel from the BS to $Q_S$ and $Q_A$ 
are approximated as $\mathbf{h}_S\approx \tilde{\mathbf{h}}_S= h_{S,\mathrm{LoS}} \mathbf{h}^{H}(r_S,\theta_S)$ and $\mathbf{h}_A\approx \tilde{\mathbf{h}}_A =h_{A,\mathrm{LoS}} \mathbf{h}^{H}(r_A,\theta_A)$, respectively, where $h_{S,\mathrm{LoS}} = \frac{\lambda}{4\pi  r_S}\exp\left(-\frac{1}{2}K(f)r_S\right)$, $h_{A,\mathrm{LoS}} = \frac{\lambda}{4\pi r_A}\exp\left(-\frac{1}{2}K(f)r_A\right)$, and $ \mathbf{h}^{H}(r_S,\theta_S)$ and $\mathbf{h}^{H}(r_A,\theta_A)$ are channel steering vectors of $Q_S$ and $Q_A$, respectively. 

For ease of implementation, we consider the maximum ratio transmission (MRT) based analog beam focusing for $Q_S$ and $Q_A$, 
i.e., $\mathbf{w_s} = \sqrt{(1-\alpha)P}\frac{\tilde{\mathbf{h}}_S}{||\tilde{\mathbf{h}}_S||}$ and $\mathbf{w_z} = \sqrt{\alpha P}\frac{\tilde{\mathbf{h}}_A}{||\tilde{\mathbf{h}}_A||}$, to facilitate beam focusing, where $\alpha\in[0,1)$ is the power allocation ratio on the AN. Given the correlation between any two near-field channel steering vectors, defined in \cite[\textbf{Definition 1}]{zhang2024performanceanalysislowcomplexitybeamforming}, the correlations between the channel steering vectors of the legitimate user and $Q_S$, and between the channel steering vectors of the legitimate user and $Q_A$ are given by
\begin{align}
\rho_1 = \rho(r_B,\theta_B,r_S,\theta_S) = |\mathbf{h}^H(r_B,\theta_B)\mathbf{h}(r_S,\theta_S)|
\end{align}
and
\begin{align}
\rho_2 = \rho(r_B,\theta_B,r_A,\theta_A) = |\mathbf{h}^H(r_B,\theta_B)\mathbf{h}(r_A,\theta_A)|,
\end{align}
respectively. Similarly, the correlations between the channel steering vectors of the eavesdropper and $Q_S$, and between the channel steering vectors of the eavesdropper and $Q_A$ are given by
\begin{align}
\rho_3 = \rho(r_E,\theta_E,r_S,\theta_S) = |\mathbf{h}^H(r_E,\theta_E)\mathbf{h}(r_S,\theta_S)|
\end{align}
and
\begin{align}
\rho_4 = \rho(r_E,\theta_E,r_A,\theta_A) = |\mathbf{h}^H(r_E,\theta_E)\mathbf{h}(r_A,\theta_A)|,
\end{align}
respectively. It is worth noting that these correlations belong to a fixed interval, i.e., $\rho_i\in(0,1]$, for $i=\{1,2,3,4\}$. Accordingly, the achievable secrecy rate for $R_S>0$ is rewritten as
\begin{align}
&R_S = R_B-R_E\notag\\
&=\log_2\left(1+\frac{|\mathbf{h}_{B}^{H} \mathbf{w_s}|^2}{|\mathbf{h}_{B}^{H}\mathbf{w_z}|^2+\sigma^2}\right) - \log_2\left(1+\frac{|\mathbf{h}_{E}^{H} \mathbf{w_s}|^2}{|\mathbf{h}_{E}^{H}\mathbf{w_z}|^2+\sigma^2}\right)\notag\\
&=\log_2\left(1+\frac{(1-\alpha)P g_B\rho_1^2}{\alpha P g_B \rho_2^2+\sigma^2}\right)-\log_2\left(1+\frac{(1-\alpha)P g_E\rho_3^2}{\alpha P g_E \rho_4^2+\sigma^2}\right)\notag\\
& = \log_2\left(\frac{A+B}{A+C}\right),
\end{align}
where $g_B = N |h_{B,\mathrm{LoS}}|^2$, $g_E = N |h_{E,\mathrm{LoS}}|^2$,
\begin{align}
A = \alpha^2 P^2\rho_2^2\rho_4^2g_Bg_E+\alpha P\sigma^2(\rho_2^2g_B+\rho_4^2g_E)+\sigma^4,
\end{align}
\begin{align}
B=(1-\alpha)P\rho_1^2g_B(\alpha P\rho_4^2g_E+\sigma^2),
\end{align}
and
\begin{align}
C = (1-\alpha)P\rho_3^2g_E(\alpha P\rho_2^2g_B+\sigma^2).
\end{align}

\subsection{Beam Focusing Design} \label{Sec:beamforcusing}

In this subsection, we determine the coordinates of both $Q_S$ and $Q_A$ for the signal and AN transmission to maximize the secrecy rate. Thus, the secrecy rate can be approximated as
\begin{align} \label{eq:RSapprox}
R_S &\overset{(a)}{\approx}\log_2\left(\frac{\alpha^2P^2\rho_2^2\rho_4^2g_Bg_E+\alpha(1-\alpha)P^2\rho_1^2\rho_4^2g_Bg_E}{\alpha^2P^2\rho_2^2\rho_4^2g_Bg_E+\alpha(1-\alpha)P^2\rho_2^2\rho_3^2g_Bg_E}\right) \notag\\
&= \log_2\left(\frac{\eta+{\rho_1^2}/{\rho_2^2}}{\eta+{\rho_3^2}/{\rho_4^2}}\right),
\end{align}
where $\eta = \frac{\alpha}{1-\alpha}\geq 0$ and $(a)$ is held by ignoring the noise power with respect to the received signal and AN power at both the legitimate user and the eavesdropper\footnote{{Since near-field THz systems typically operate in moderate to high SNR regions due to high beam focusing gains, both the received signal power at the legitimate user and the AN power at the eavesdropper significantly exceed the noise power. Under these conditions, we neglect the noise power to facilitate the derivation of tractable closed-form expressions.}}. 
{To ensure reliable communication at the legitimate user while still disrupting the eavesdropper, the signal transmission should dominate the total transmit power, i.e., $\eta$ should be small \cite{zhang2024performanceanalysislowcomplexitybeamforming}. Moreover, to maximize the secrecy rate, it is essential for the received signal power at the legitimate user to exceed that at the eavesdropper, i.e., the received signal power ratio $\rho_1^2/\rho_2^2$ is large and much greater than $\eta$.} 
Based on it, the secrecy rate maximization problem can be approximated as
\begin{align}
    \max\limits_{L_S,L_A}\ \log_2\left(\frac{\rho_1^2}{\rho_3^2}\right)+ \log_2\left(\frac{\rho_4^2}{\rho_2^2}\right),
\end{align}
for a small $\eta$. We note that $\rho_1$ and $\rho_3$ depend solely on the coordinate of $Q_S$, $L_S$, while $\rho_2$ and $\rho_4$ depend solely on the coordinate of $Q_A$, $L_A$. Thus, the original optimization problem $\mathbf{P1}$ can be approximately decoupled to
\begin{align}
\mathbf{P2}:\quad &\max\limits_{{L_S}}\  \log_2\left(\frac{\rho_1^2}{\rho_3^2}\right),
\end{align}
which maximizes the received signal power ratio between the legitimate user and the eavesdropper, and
\begin{align}
\mathbf{P3}:\quad &\max\limits_{{L_A}}\  \log_2\left(\frac{\rho_4^2}{\rho_2^2}\right),
\end{align}
which maximizes the received AN power ratio between the eavesdropper and the legitimate user. {We note that the decoupled sub-problems $\mathbf{P2}$ and $\mathbf{P3}$ are independent of each other, and both are independent of the power allocation ratio, $\alpha$.} Since the received signal power at the legitimate user is maximized when $\theta_S = \theta_B$, and the received AN power at the eavesdropper is maximized when $\theta_A = \theta_E$ \cite[\textbf{Proposition 1}]{zhang2024performanceanalysislowcomplexitybeamforming}, we adopt $\theta_S = \theta_B$ and $\theta_A = \theta_E$, and need to determine $r_S$ and $r_A$ to find the optimal $Q_S$ and $Q_A$. 
With $\theta_S = \theta_B$ and \cite[\textbf{Lemma 1}]{Chen2023wcl}, $\rho_1$ and $\rho_3$ can be approximated as
\begin{align}
\rho_1 \approx \left|\frac{C(\beta_1)+j S(\beta_1)}{\beta_1}\right|
\end{align}
and
\begin{align}
\rho_3 \approx \left|\frac{\tilde{C}(\beta_2,\beta_3)+j \tilde{S}(\beta_2,\beta_3)}{2\beta_3}\right|,
\end{align}
respectively, where $C(\beta) = \int_0^{\beta}\cos\left(\frac{\pi}{2}t^2\right)\mathrm{d}t$ and $S(\beta) = \int_0^{\beta}\sin\left(\frac{\pi}{2}t^2\right)\mathrm{d}t$ are Fresnel integrals, $\tilde{C}(\beta_2,\beta_3) = C(\beta_2+\beta_3)-C(\beta_2-\beta_3)$, $\tilde{S}(\beta_2,\beta_3) = S(\beta_2+\beta_3)-S(\beta_2-\beta_3)$, with $\beta_1 = \frac{N}{2}\sqrt{d\left|\frac{1-\theta_B^2}{r_B}-\frac{1-\theta_S^2}{r_S}\right|}$, $\beta_2 = \left(\theta_E-\theta_S\right)/\sqrt{d\left|\frac{1-\theta_E^2}{r_E}-\frac{1-\theta_S^2}{r_S}\right|}$, and $\beta_3 = \frac{N}{2}\sqrt{d\left|\frac{1-\theta_E^2}{r_E}-\frac{1-\theta_S^2}{r_S}\right|}$.
Thus, the near-optimal $r_{S,\textrm{Opt}}$ is the solution to
\begin{align}
\max_{{r_S}}\left(\frac{\beta_3^2(C^2(\beta_1)+S^2(\beta_1))}{\beta_1^2(\tilde{C}^2(\beta_2,\beta_3)+ \tilde{S}^2(\beta_2,\beta_3))}\right),
\end{align}
which can be solved using the one-dimensional search method. {The near-optimal $r_{A,\textrm{Opt}}$ can also be obtained using an approach similar to that used for determining $r_{S,\textrm{Opt}}$.}

\subsection{Power Allocation Design}

We next determine the optimal power allocation ratio, $\alpha$, in this subsection. We note that the signal and AN beam focusing design in Sec. \ref{Sec:beamforcusing} achieves the near-optimal secrecy performance by ignoring the noise power, regardless of $\alpha$. We now examine the effect of this noise power on the optimal power allocation ratio in our design.

\Lemma\label{Lemma:1} The optimal power allocation ratio to achieve the maximum secrecy rate is obtained as
\begin{align}
\alpha = \left\{
\begin{aligned}
&\frac{-F_1-\sqrt{F_1^2-4F_2F_0}}{2F_2},&\textrm{if } F_0>0 \\
&0, &\textrm{otherwise,}
\end{aligned}
\right.
\end{align} 
where $F_0$, $F_1$, and $F_2$ are given by
\begin{align}
F_0=&P\sigma^2(P^2g_Bg_E\rho_1^2\rho_3^2(g_E\rho_4^2-g_B\rho_2^2)\notag\\
&+P\sigma^2(\rho_3^2\rho_4^2g_E^2-\rho_1^2\rho_2^2g_B^2)+\sigma^4(\rho_3^2g_E-\rho_1^2g_B)),
\end{align}
\begin{align}
F_1=&2P^2g_Bg_E\sigma^2(Pg_B\rho_1^2\rho_2^2(\rho_3^2-\rho_4^2)\notag\\
&+Pg_E\rho_3^2\rho_4^2(\rho_2^2-\rho_1^2)+\sigma^2(\rho_2^2\rho_3^2-\rho_1^2\rho_4^2)),
\end{align}
and
\begin{align}
F_2=&P^3 g_B g_E (Pg_Bg_E\rho_2^2\rho_4^2(\rho_2^2\rho_3^2-\rho_1^2\rho_4^2) \notag\\
&+g_B\rho_2^2\rho_3^2\sigma^2(\rho_2^2-\rho_1^2)+g_E\rho_1^2\rho_4^2\sigma^2(\rho_3^2-\rho_4^2)),
\end{align}
respectively.

\begin{IEEEproof}
Here, we define $\Omega=\frac{A+B}{A+C}$ and note that the maximum secrecy rate is achieved when $\Omega$ achieves its maximum value. To find the maximum value of $\Omega$, we first evaluate the first derivative of $\Omega$ with respect to $\alpha$ as
\begin{align}\label{eq:firstOrder}
\frac{\partial\Omega }{\partial \alpha} = \frac{F_2\alpha^2+F_1\alpha+F_0}{(A+C)^2}.     
\end{align}
Since the beam focusing design in Sec. \ref{Sec:beamforcusing} maximizes $\frac{\rho_1^2}{\rho_2^2}$ and $\frac{\rho_4^2}{\rho_3^2}$, it can be easily verified that $F_2$ and $F_1$ are always smaller than zero. Therefore, the monotonicity of $\Omega$ relies on the sign of $F_0$. On the one hand, if $F_0>0$, $\Omega$ first increases and then decreases as $\alpha$ increases from $0$ to $1$. Thus, the maximum secrecy rate is achieved when $\frac{\partial\Omega }{\partial \alpha}=0$. On the other hand, when $F_0\leq 0$, $\frac{\partial\Omega }{\partial \alpha}$ is always smaller than zero, thereby resulting in a monotonically decreasing secrecy rate with respect to $\alpha$.
\end{IEEEproof}

\textbf{Remark 1:} We note that the AN helps improve the secrecy rate only when $F_0>0$ and the sign of $F_0$ primarily depends on the last term, i.e., $\rho_3^2g_E-\rho_1^2g_B$. Since $g_E$ and $g_B$ decrease when $r_E$ and $r_B$ increase, respectively, we find that the injection of the AN into signal transmission can improve the secrecy rate when the eavesdropper is located closer to the BS than the legitimate user, i.e., $g_E>g_B$. Otherwise, the improvement in the secrecy rate by the AN is limited. This will be illustrated in Sec. \ref{Sec:Num}.

Finally, our proposed AN aided beam focusing scheme can be obtained by combining the signal and AN beam focusing design in Sec. \ref{Sec:beamforcusing} with the power allocation design in Lemma \ref{Lemma:1}. 
Its benefits will be examined in Sec. \ref{Sec:Num}.

\section{Numerical Results}\label{Sec:Num}

In this section, we present numerical results to demonstrate the effectiveness of our proposed AN aided beam focusing scheme in Sec. \ref{Sec:Coverage}. We set $N=513$ antennas in the large-scale array, carrier frequency as $f=300$ GHz, antenna spacing as $d=\lambda/2$, noise power as $\sigma^2 = -77$ dBm for a $5$ GHz bandwidth, based on existing experimental platforms \cite{SEN2020107370}, transmit power as $P = 5$ dBm, the spatial angle from the x-axis to the legitimate user and the eavesdropper as $\theta_B=\theta_E=0$, and the molecular absorption coefficient at frequency $f$ as $K(f)=0.00143$ $\mathrm{m}^{-1}$. {Moreover, the number of NLoS paths from the BS to the legitimate user and eavesdropper is set as $M_B=M_E = 2$, where each NLoS path has an average path loss coefficient that is $15$ dB lower than that of the LoS path \cite{zhang2024performanceanalysislowcomplexitybeamforming}.} {For comparison, we employ the state-of-the-art AN aided beam focusing scheme proposed in \cite{zhang2024performanceanalysislowcomplexitybeamforming} as the benchmark scheme.}

\begin{figure}[t]
    \centering
        \subfigure[Signal beam focusing with $r_E=3.5$ m.]
        {    \label{fig:BeamFS35}        
            \includegraphics[width=0.45\columnwidth]{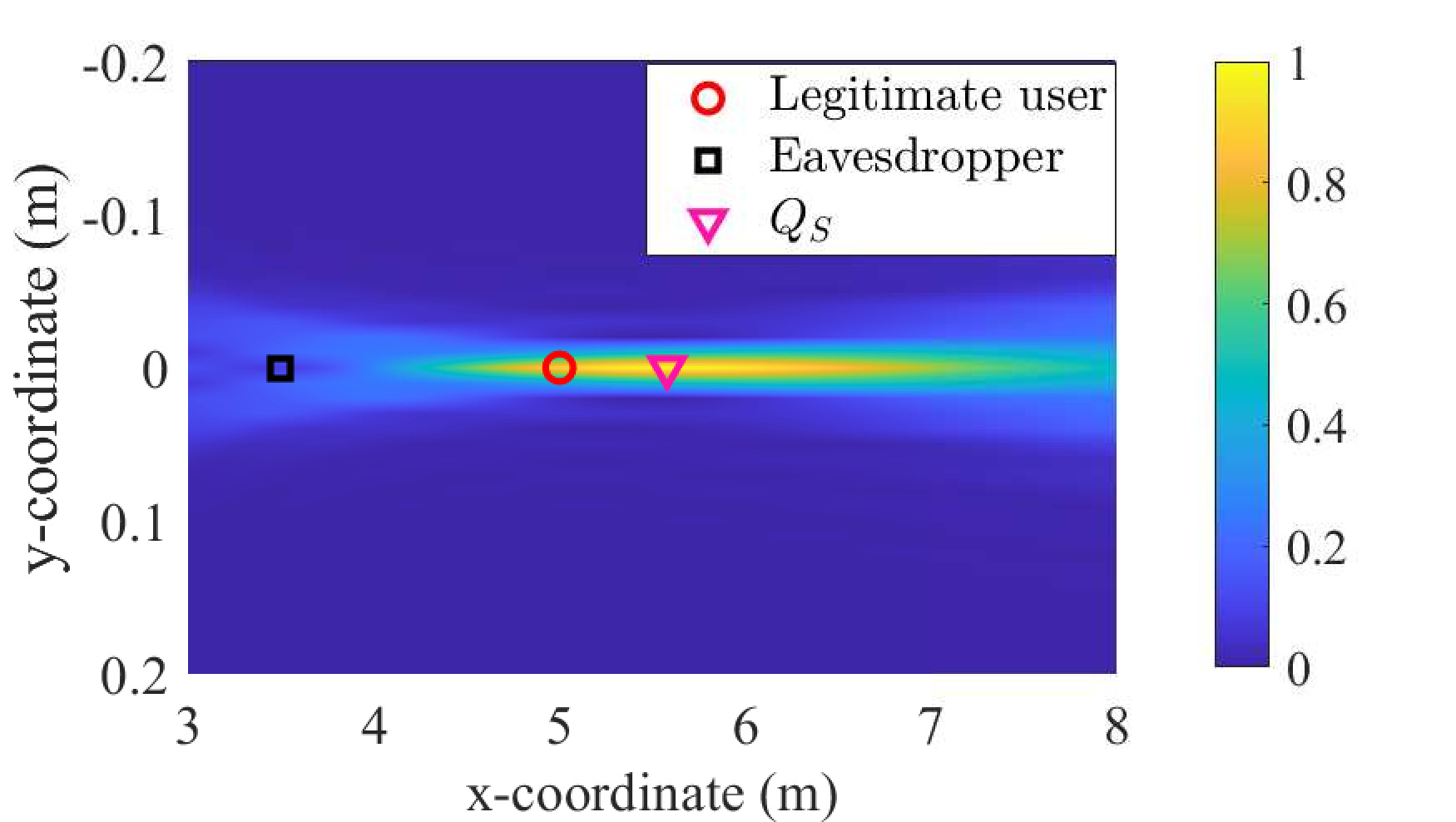}
            }
        \subfigure[AN beam focusing with $r_E=3.5$ m.]{\label{fig:BeamFA35}
             \includegraphics[width=0.45\columnwidth]{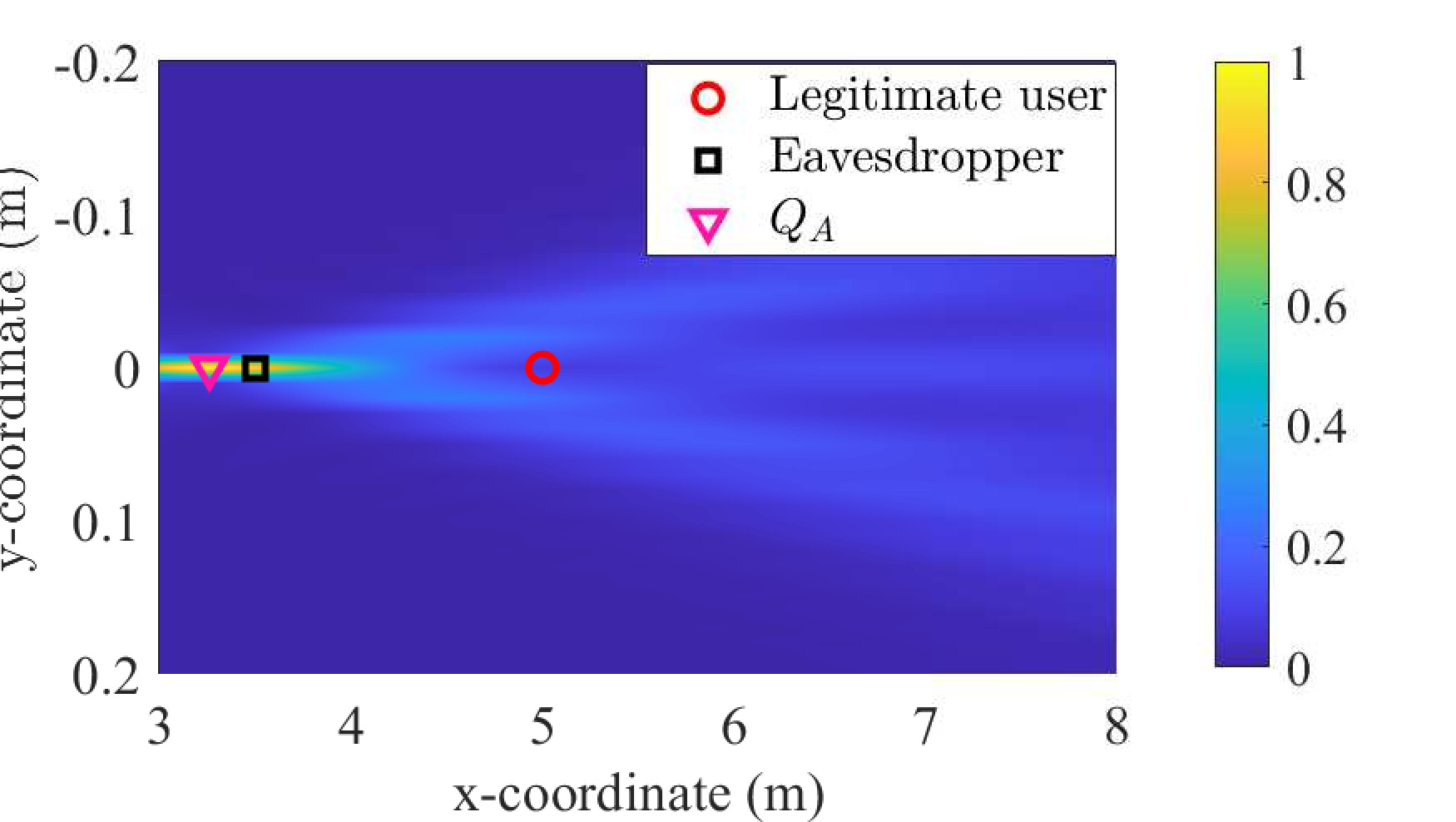}
            }
        \subfigure[Signal beam focusing with $r_E=4$ m.]
        {    \label{fig:BeamFS4}       
            \includegraphics[width=0.45\columnwidth]{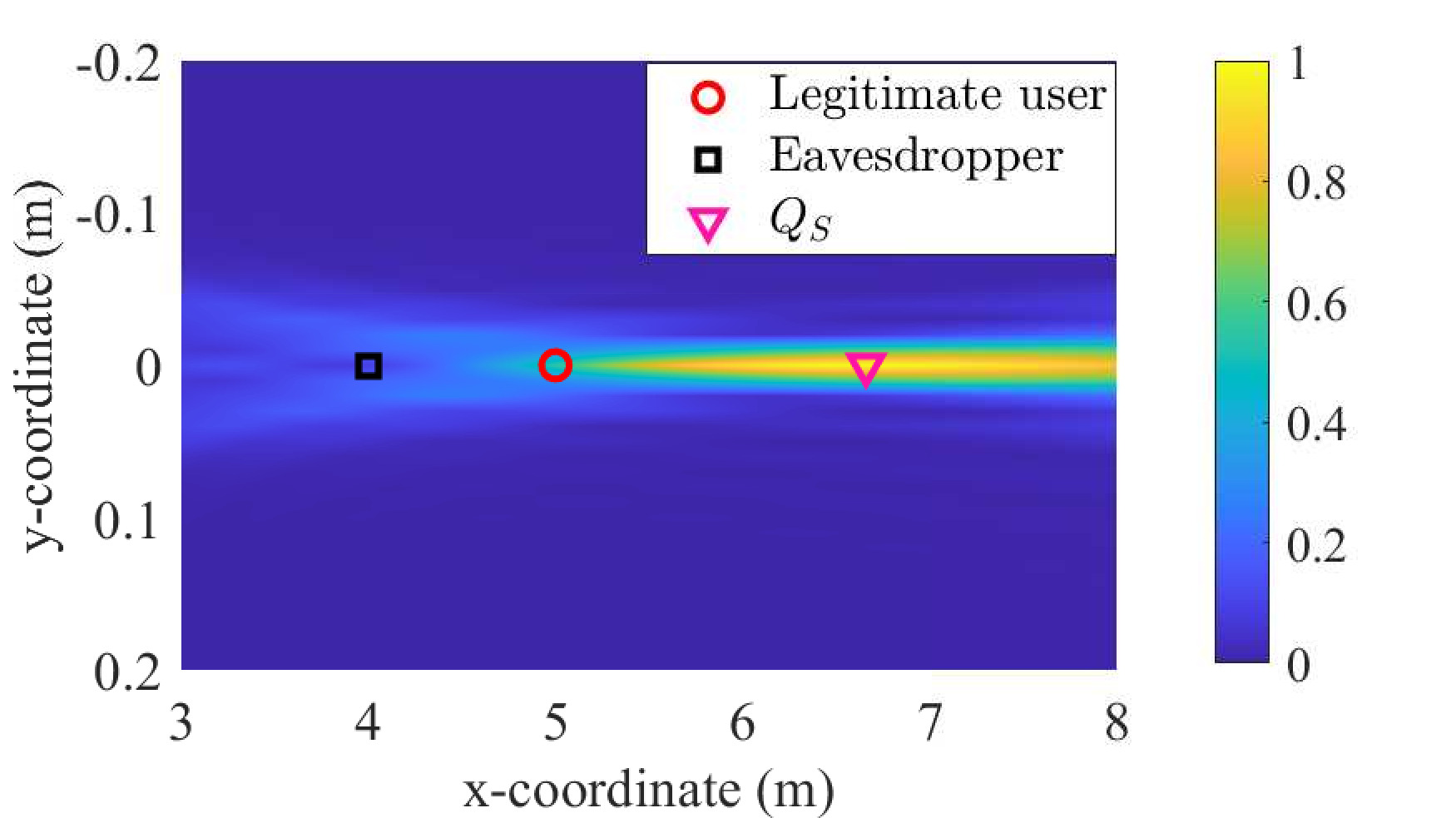}
            }
        \subfigure[AN beam focusing with $r_E=4$ m.]{\label{fig:BeamFA4}
             \includegraphics[width=0.45\columnwidth]{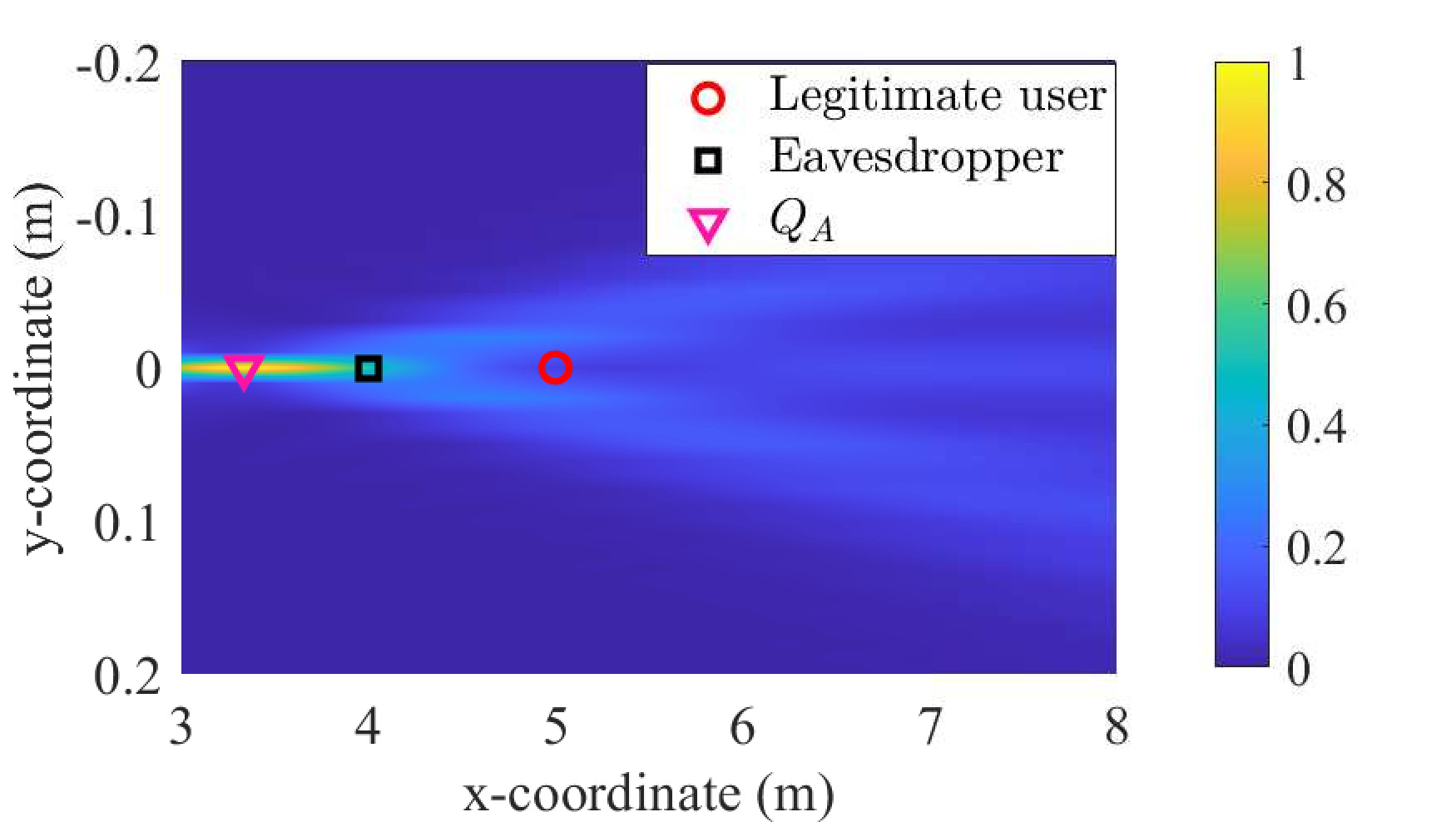}
            }\vspace{-0.5em}
             \subfigure[Signal beam focusing with $r_E=4$ m and $\theta_E = \frac{\pi}{1800}$.]
        {    \label{fig:BeamFS4_angle}       
            \includegraphics[width=0.45\columnwidth]{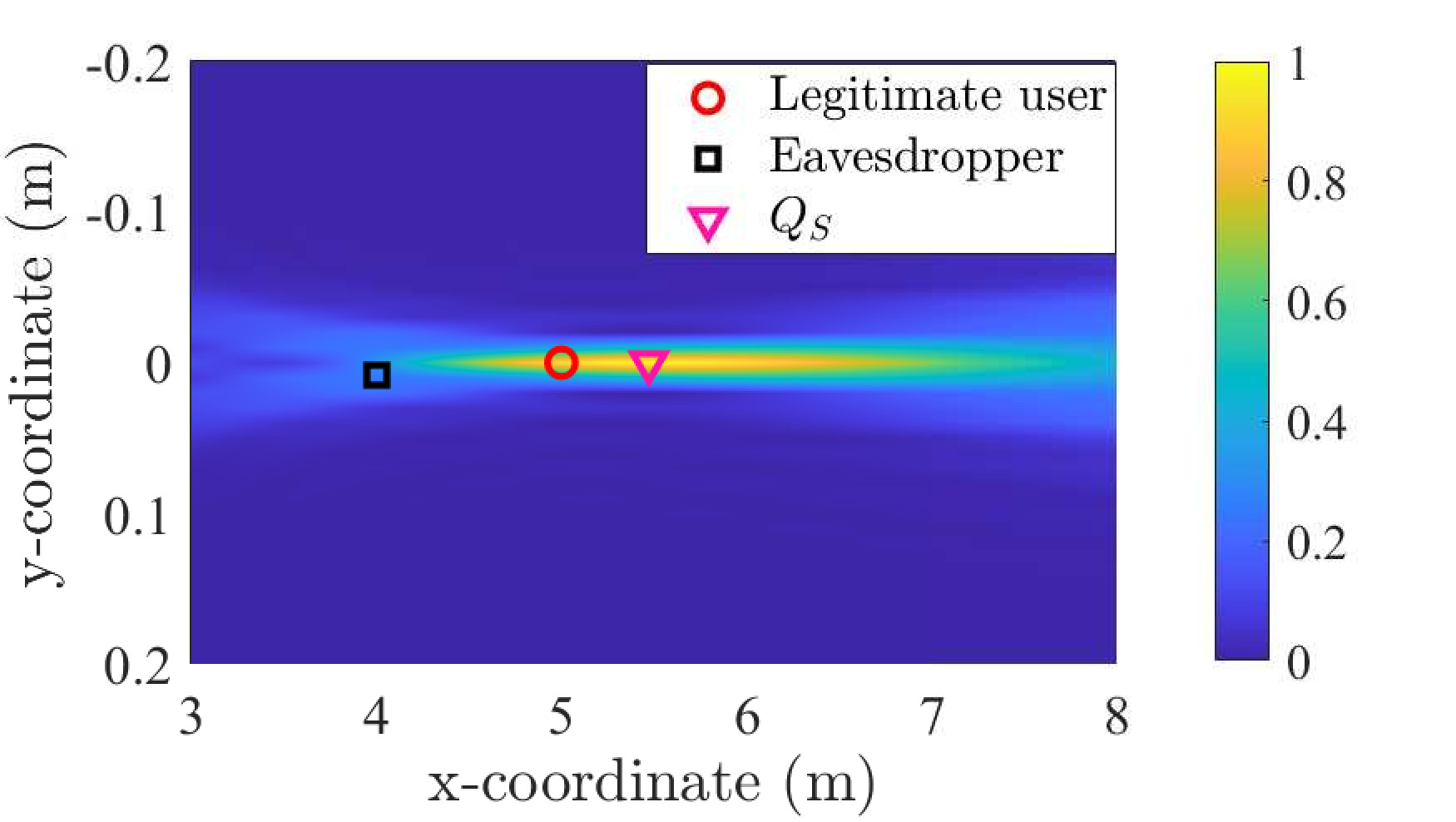}
            }
        \subfigure[AN beam focusing with $r_E=4$ m and $\theta_E = \frac{\pi}{1800}$.]{\label{fig:BeamFA4_angle}
             \includegraphics[width=0.45\columnwidth]{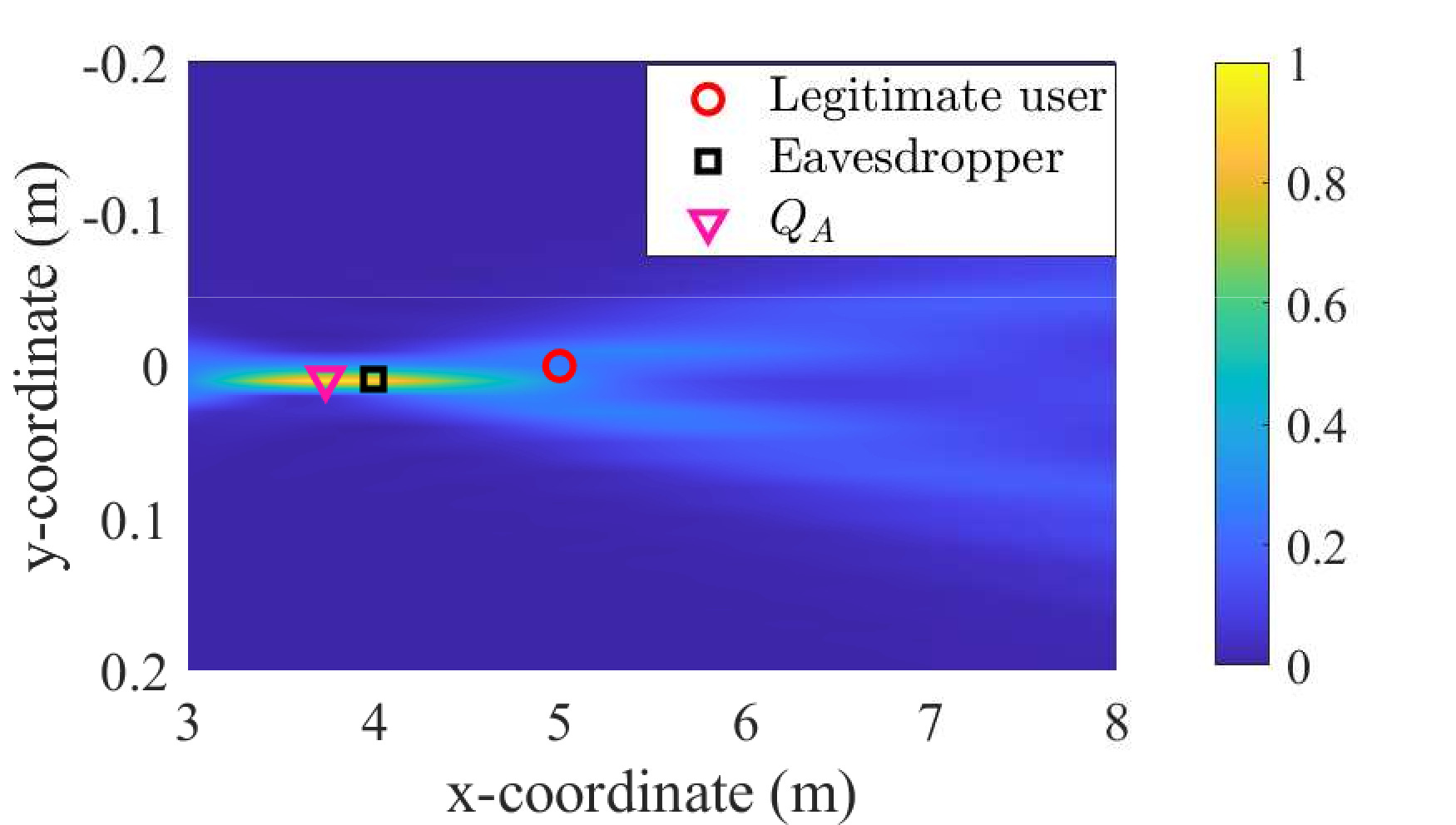}
            }\vspace{-0.5em}
        \caption{The normalized signal power spectrum of the signal and AN beam focusing design with $r_B=5$ m.}
        \vspace{-1em}
        \label{fig:Beam}
\end{figure}

Fig.~\ref{fig:Beam} plots the normalized signal power spectrum of the proposed signal and AN beam focusing design with $r_B=5$ m and different $r_E$ {and $\theta_E$}. We observe that the proposed $Q_S$ and $Q_A$ are not allocated to the legitimate user and the eavesdropper, respectively. {Moreover, the distance between the legitimate user and the proposed $Q_S$, and that between the eavesdropper and the proposed $Q_A$, decrease as the eavesdropper moves closer to the legitimate user.} This observation is due to the two-fold impact on the secrecy rate when the distance from $Q_S$ to the legitimate user and that from $Q_A$ to the eavesdropper decrease. Specifically, the reduction in such distances increases both the received signal power at the legitimate user and the received AN power at the eavesdropper, as well as boosts the received signal power at the eavesdropper and the received AN power at the legitimate user, especially when the eavesdropper is close to the legitimate user. Importantly, this two-fold impact is addressed in our proposed beam focusing design. {In addition, we observe that the distance between the legitimate user and the proposed $Q_S$, and that between the eavesdropper and the proposed $Q_A$, increase as the spatial angle difference between the eavesdropper and legitimate user increases. This observation is due to the fact that the increase in the spatial angle difference between the legitimate user and eavesdropper decreases the correlation between their channel steering vectors, allowing $Q_S$ to be designed closer to the legitimate user to improve the received signal power at the legitimate user, while $Q_A$ can be positioned closer to the eavesdropper to enhance the received AN at the eavesdropper.}

\begin{figure}
    \centering
    \includegraphics[width=0.78\columnwidth]{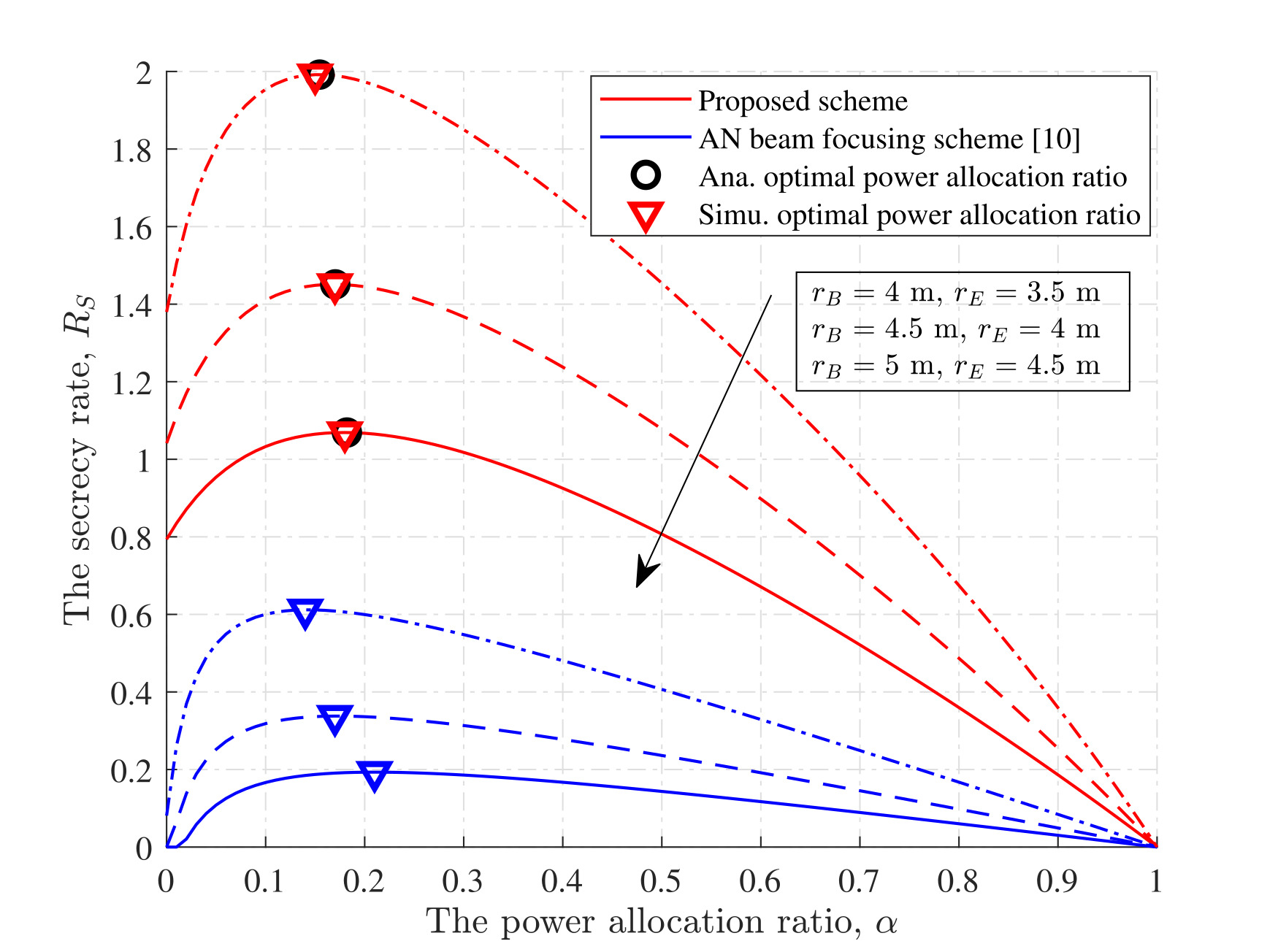}
    \vspace{-1em}
    \caption{The secrecy rate, $R_S$, versus the power allocation ratio, $\alpha$.}
    \vspace{-1.2em}
    \label{fig:num1}
\end{figure}

Fig.~\ref{fig:num1} plots the secrecy rate, $R_S$, versus the power allocation ratio, $\alpha$. We first observe that our proposed scheme achieves a higher secrecy rate than the benchmark scheme, which shows the superiority of our proposed scheme. We then observe that the optimal power allocation ratio derived in Lemma \ref{Lemma:1} achieves the optimal secrecy rate, demonstrating the accuracy of our analysis. We further observe that $R_S$ first increases and then decreases as $\alpha$ increases. This observation is due to the fact that the increase in $\alpha$ has a two-fold impact on the secrecy rate. Specifically, when $\alpha$ is small, its increase leads to a higher received AN power at the eavesdropper, thereby decreasing the channel capacity of the eavesdropper. However, once $\alpha$ exceeds a certain threshold, its increase leads to a profound decrease in the received signal power at the legitimate user, resulting in a lower channel capacity at the legitimate user and consequently the decreased secrecy rate.

\begin{figure}
    \centering
    \includegraphics[width=0.78\columnwidth]{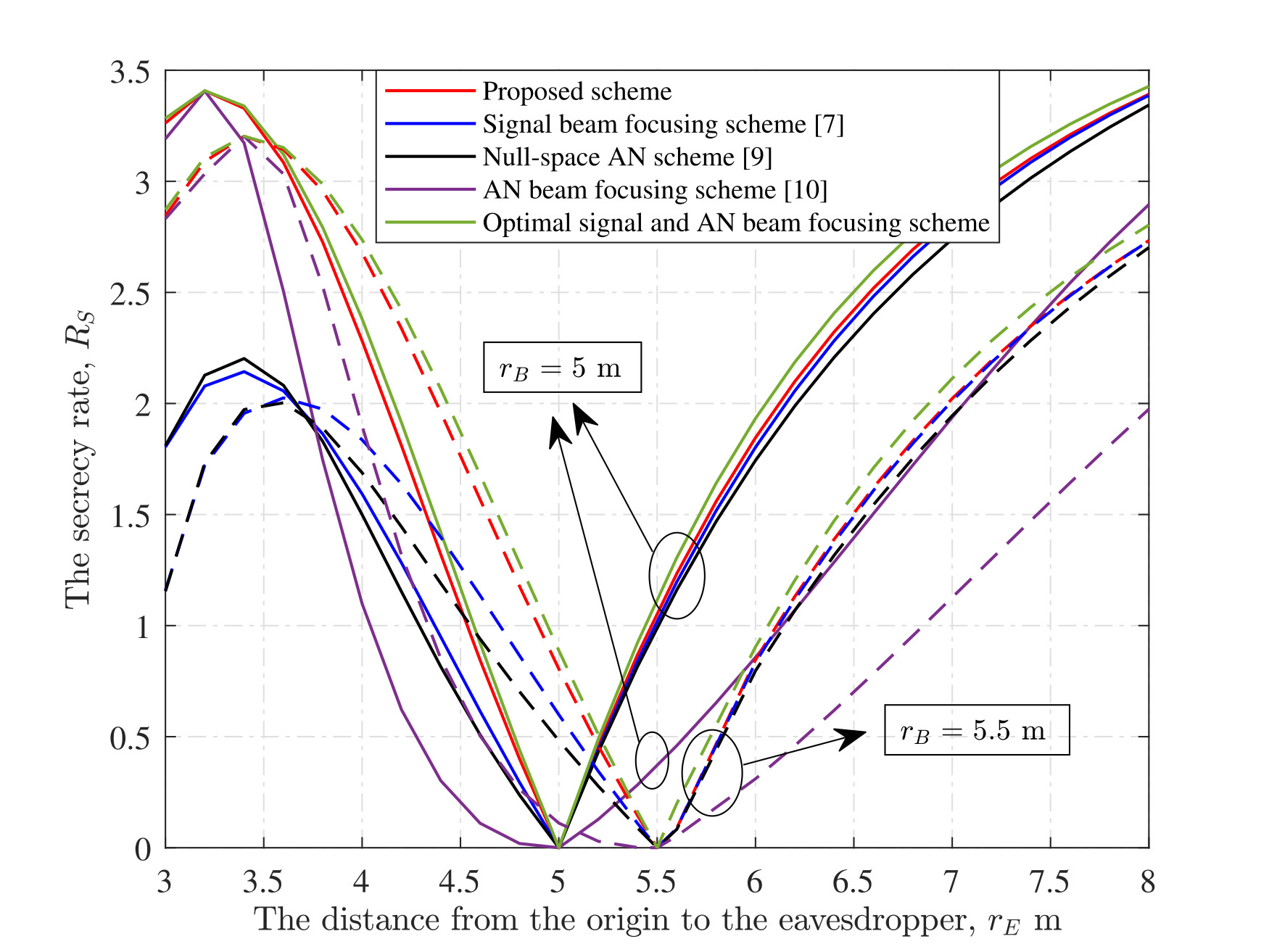}
    \vspace{-1em}
    \caption{The secrecy rate, $R_S$, versus the distance from the origin to the eavesdropper, $r_E$, with different $r_B$.}
    \vspace{-1.2em}
    \label{fig:num2}
\end{figure}

Fig.~\ref{fig:num2} plots the secrecy rate, $R_S$, versus the distance from the origin to the eavesdropper, $r_E$, with $r_B = 5$ m. {We compare our proposed scheme with the state-of-the-art signal beam focusing scheme proposed in \cite{Ali2024coml}, null-space AN aided scheme proposed in \cite{Chen2024tvt}, and AN aided beam focusing scheme proposed in \cite{zhang2024performanceanalysislowcomplexitybeamforming} as the benchmark schemes. Specifically, \cite{Ali2024coml} employs signal beam focusing at the legitimate user, \cite{Chen2024tvt} projects AN onto the null space of the legitimate channel, and \cite{zhang2024performanceanalysislowcomplexitybeamforming} applies AN beam focusing toward the eavesdropper.} 
We first observe that for all the schemes, the secrecy rate decreases and then increases as the distance from the origin to the eavesdropper, $r_E$, increases. This observation is due to the fact that the channels from the BS to the legitimate user and the eavesdropper are more similar when the eavesdropper is located close to the legitimate user, where this similarity significantly reduces the secrecy rate in near-field communication systems. {We then observe that the performance of our proposed scheme is very close to the optimal signal and AN beam focusing scheme. Moreover, it consistently outperforms the benchmark schemes by achieving a higher secrecy rate for different values of $r_B$. }
Compared to the scheme in \cite{Ali2024coml}, we find that our proposed scheme provides significant secrecy rate improvement when the eavesdropper is located closer to the BS than the legitimate user, i.e., $r_E<r_B$, while this improvement is not notable when the eavesdropper is located farther to the BS than the legitimate user, i.e., $r_E>r_B$. This observation is in accordance with Remark 1, which further validates our analysis. We clarify that the results for the optimal signal and AN beam focusing scheme are obtained by solving $\mathbf{P1}$, incurring the computational complexity $\mathcal{O}((NM)^3)$, where $N$ is the number of antennas and $M$ is a constant determined by the desired precision requirement. {The schemes in \cite{Ali2024coml}, \cite{Chen2024tvt}, and \cite{zhang2024performanceanalysislowcomplexitybeamforming} have computational complexities of $\mathcal{O}(N^3)$, $\mathcal{O}(N^2)$, and $\mathcal{O}(N)$, respectively, while the computational complexity of our proposed scheme is $\mathcal{O}(NM)$.} Thus, our proposed scheme is computationally efficient compared to the benchmark schemes. 

\section{Conclusion}\label{Sec:Conclusion}

We proposed a novel design to safeguard information transmission for a single legitimate user against the risk of eavesdropping in a near-field THz communication system. By decoupling the original optimization problem, we solved each subproblem with a cost-cautious one-dimensional search method and derived the closed-form expression for the optimal power allocation ratio. We then combined the solutions to the two decoupled subproblems and devised a low-complexity AN aided beam focusing scheme to maximize the secrecy rate. Aided by numerical results, we demonstrated a significant performance gain achieved by the proposed scheme compared to the benchmark scheme, as well as discussed the insights arising from our proposed AN aided design.

\bibliographystyle{IEEEtran} 
\bibliography{bibli}

\end{document}